\documentclass[10pt,twocolumn,letterpaper]{article}

\usepackage[pagenumbers]{cvpr} %

\usepackage{pifont}%

\newcommand{\rom}[1]{\uppercase\expandafter{\romannumeral #1\relax}}

\usepackage{calc}

\usepackage{amsthm}
\usepackage{amssymb}
\usepackage[dvipsnames]{xcolor}
\definecolor{cvprblue}{rgb}{0.21,0.49,0.74}
\usepackage[pagebackref,breaklinks,colorlinks,citecolor=cvprblue]{hyperref}
\usepackage{colortbl}
\usepackage{tabulary}
\usepackage{tabularx}
\usepackage{etoolbox}
\usepackage{multirow}
\usepackage{cuted}
\usepackage[percentage]{overpic}
\usepackage{booktabs}
\usepackage{pgfplots}

\pgfplotsset{compat=newest}%

\def\paperID{173} %
\def\confName{3DV\xspace}
\def\confYear{2026\xspace}

\title{Hyper Diffusion Avatars: Dynamic Human Avatar Generation using Network Weight Space Diffusion}

\author{Dongliang Cao\\
University of Bonn\\
\and
Guoxing Sun\\
Max Planck Institute for Informatics\\
\and
Marc Habermann\\
Max Planck Institute for Informatics\\
\and
Florian Bernard\\
University of Bonn\\
}

\begin{document}
\maketitle
\begin{strip}
  \centerline{
  \vspace{-5mm}
  \footnotesize
  \begin{tabular}{c}
    \setlength{\tabcolsep}{0pt}
    \includegraphics[width=\textwidth]{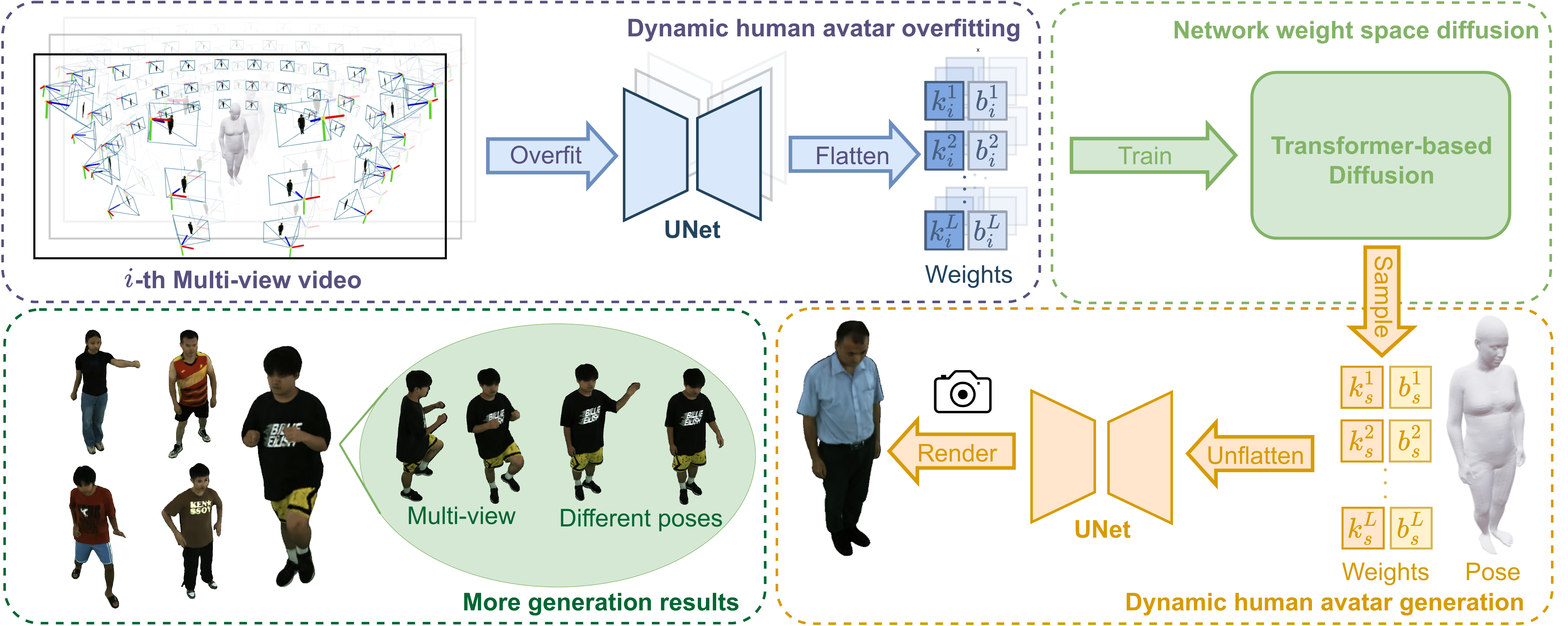}
  \end{tabular}
  }
\captionof{figure}{
\textbf{Our method enables dynamic human avatar generation via diffusion in network weight space.} 
First, we optimize a set of UNets, each representing an individual dynamic human avatar (top left). 
Next, we train a transformer network to model a diffusion process over these optimized network weights (top right). 
At inference time, our approach samples new network weights for real-time, controllable dynamic human avatar rendering by predicting pose-dependent 3D Gaussian Splatting based on a given pose (bottom).
}
\label{fig:teaser}
\vspace{-3mm}
\end{strip}

\begin{abstract}  
Creating human avatars is a highly desirable yet challenging task. Recent advancements in radiance field rendering have achieved unprecedented photorealism and real-time performance for personalized dynamic human avatars. However, these approaches are typically limited to person-specific rendering models trained on multi-view video data for a single individual, limiting their ability to generalize across different identities. On the other hand, generative approaches leveraging prior knowledge from pre-trained 2D diffusion models can produce cartoonish, static human avatars, which are animated through simple skeleton-based articulation. Therefore, the avatars generated by these methods suffer from lower rendering quality compared to person-specific rendering methods and fail to capture pose-dependent deformations such as cloth wrinkles. 
In this paper, we propose a novel approach that unites the strengths of person-specific rendering and diffusion-based generative modeling to enable \textbf{dynamic human avatar} generation with both high photorealism and realistic pose-dependent deformations. Our method follows a two-stage pipeline: first, we optimize a set of person-specific UNets, with each network representing a dynamic human avatar that captures intricate pose-dependent deformations. In the second stage, we train a hyper diffusion model over the optimized network weights. During inference, our method generates network weights for real-time, controllable rendering of dynamic human avatars. Using a large-scale, cross-identity, multi-view video dataset, we demonstrate that our approach outperforms state-of-the-art human avatar generation methods.
\end{abstract}
    
\section{Introduction}
\label{sec:intro}

Generating high-quality renderings of humans is a crucial challenge in computer vision and computer graphics, with numerous real-world applications in remote communication, movies, gaming, and immersive experiences in augmented as well as virtual reality. 
Traditionally, generating digital avatars from real-world data requires complicated hardware setups, manual efforts from skilled artists, and advanced physical-based rendering techniques to synthesize the final image~\cite{bickel2007multi,garrido2013reconstructing}. 
\par
With the advancement of neural radiance fields and subsequent works~\cite{mildenhall2021nerf, wang2021neus, wang2023neus2, kerbl20233d}, recent methods~\cite{liu2021neural, habermann2023hdhumans, kwon2023deliffas, zhu2024trihuman,pang2024ash} have focused on learning photorealistic and controllable human avatars directly from calibrated multi-view videos. 
Although these approaches achieve unprecedented levels of photorealism, they are still person-specific, meaning that for each individual human a dense multi-view video has to be captured, data has to be processed and annotated, and a dedicated neural model has to be trained from scratch. 
This process is neither scalable nor fast and resource-efficient as these steps can easily take multiple days~\cite{liu2021neural, habermann2023hdhumans}.
\par
Meanwhile, recent generative methods~\cite{rombach2022high,podell2023sdxl,esser2024scaling} have made significant progress in generalization quality and scalability, driven by advances in generative diffusion models~\cite{song2020denoising,ho2020denoising,lipman2022flow}. 
To this end, recent avatar generation methods~\cite{kolotouros2023dreamhuman, liu2024humangaussian, cao2024dreamavatar, huang2024tech, liao2024tada} attempted to distill prior knowledge from 2D image diffusion models through score distillation sampling~\cite{poole2022dreamfusion}. 
Despite their compelling results, the rendering quality remains significantly lower than that of person-specific rendering methods.
Notably, their rendered videos are unable to capture skeletal pose-dependent deformations like clothing wrinkles and appearance variations, e.g. cast shadows, due to the limitations of simple skeleton-based articulation. 
To address the limitations mentioned above, for {the first time} we aim to unify the person-specific rendering method and the diffusion-based generation model to generate photorealistic real-time renderings across different individuals, which faithfully captures pose-dependent deformations.  
\par 
To this end, we represent the digital human as 3D Gaussians~\cite{kerbl20233d} that are parameterized in UV space~\cite{pang2024ash,saito2024relightable,hu2024gaussianavatar,teotia2024gaussianheads}. 
In contrast to person-specific rendering methods relying on individual mesh templates~\cite{pang2024ash,li2024animatable}, we use a parametric human body model (i.e.\ SMPL-X~\cite{loper2015smpl,pavlakos2019smplx}) to offer a canonical template and a consistent UV space across individuals~\cite{zhang20243gen}. 
However, instead of directly optimizing the 3D Gaussian parameters defined in UV space for each individual, we optimize a UNet~\cite{ronneberger2015u} that maps the human pose into the Gaussian parameters defined in UV space. 
To this end, our method is capable of capturing pose-dependent deformation by predicting motion-aware 3D Gaussian parameters. 
After optimizing the person-specific network for all individuals, we propose a hyper diffusion model, which generates network weights of the optimized UNet rather than 3D Gaussian parameters directly. 
The motivation for training a diffusion model in this network weight space is two fold: 
(1) the single network encodes comprehensive pose-dependent 3D Gaussian parameters, as opposed to a static UV Gaussian map; 
(2) the network weights provide a shared canonical representation across different individuals, as opposed to person-specific rendering methods. 
During inference, we can directly use our diffusion model to sample network weights and use the generated network to render dynamic digital avatars with the skeletal pose as the input. 
Fig.~\ref{fig:teaser} provides an overview of our method. 
We summarize our main contributions as follows:
\begin{itemize}
    \item For the first time, we unify person-specific rendering and diffusion-based generation to enable dynamic human avatar generation with \emph{pose-dependent deformations}.
    \item To this end, we encode a dynamic human avatar into a motion-aware network and learn a hyper diffusion model that generates the network weights representing a dynamic avatar.
    \item To train our hyper diffusion model on network weights, we leverage a transformer-based diffusion model that effectively learns the complex structure of these weights.
\end{itemize}

\section{Related work} \label{sec:related_work}
In this work, we focus on unconditional dynamic human avatar generation. As a result, human reconstruction methods that rely on multi-view images and the corresponding human pose as inputs at inference time~\cite{remelli2022drivable, Shetty_2024_CVPR, zheng2024gps, sun2025real, zubekhin2025giga, RoGSplat2025CVPR} are out of the scope of our paper.
\subsection{Personalized 3D human rendering}
Recent advancements in neural rendering, such as NeRF~\cite{mildenhall2021nerf} and 3DGS~\cite{kerbl20233d}, have made it possible to learn human avatars directly from calibrated multi-view video inputs. 
Starting from NeRF~\cite{mildenhall2021nerf}, various approaches have been proposed to reconstruct the dynamic appearance of 3D humans~\cite{bergman2022generative, feng2022capturing, hu2023sherf, li2022tava, weng2022humannerf, liu2021neural}. 
The key idea behind these methods is to introduce deformable human NeRFs that deform the posed space to a shared pose canonical space. 
Despite producing high-quality renderings, these methods inherit the limitations of NeRF-based approaches, resulting in significantly longer rendering times. 
To overcome this limitation, more recent methods~\cite{pang2024ash, saito2024relightable, hu2024gaussianavatar, zheng2024gps, hu2024gauhuman} replace NeRF by 3DGS to enable real-time rendering speed while also improving photorealism. 
Nevertheless, the aforementioned methods primarily focus on achieving photorealistic renderings of a \emph{single personalized human}. 
In contrast, our method aims at building a generative and dynamic 3D human avatar model by training on large, cross-identity, and multi-view datasets.
\subsection{3D human generation}
Recent diffusion-based image generation models have demonstrated unprecedented progress in the context of quality, diversity, and controllability~\cite{rombach2022high, podell2023sdxl, esser2024scaling, meng2023distillation, sauer2024adversarial}. 
To this end, numerous efforts~\cite{qian2023magic123, shi2023zero123++, xu2023neurallift,tang2023make,metzer2023latent} have been made to leverage the rich 2D prior knowledge for 3D generation through score distillation sampling~\cite{poole2022dreamfusion,wang2023score}. 
Similarly, recent 3D human generation methods~\cite{liu2024humangaussian, cao2024dreamavatar, liao2024tada, huang2024tech, zhang2024humanref} also utilize the idea to optimize the underlying 3D representation, i.e. NeRF or 3DGS, given text or image conditions. 
Despite their compelling results, these methods suffer from computational inefficiency, due to the involved per-instance optimization~\cite{long2024wonder3d}. 
To improve efficiency, most recent works~\cite{chen2023primdiffusion,zhang20243gen} directly train a diffusion model in the underlying 3D representation space, e.g.\ volumetric primitives~\cite{lombardi2021mixture} or 3DGS UV maps from multi-view human data~\cite{renderpeople,tao2021function4d}. 
Nevertheless, they fail to model pose-dependent deformations by learning a static representation and using solely simple skeleton-based articulation, i.e.~linear blend skinning~\cite{lewis2023pose}. 
In contrast, our method trains a diffusion model directly in the network weight space, where the network captures pose-dependent deformations while also achieving real-time rendering speed once the weights have been initialized.
\subsection{Diffusion models for generative 3D Gaussian Splatting}
In comparison to 2D image generation, generating 3D objects is much more difficult due to the additional dimension and the scarcity of high-quality 3D data~\cite{zhang20233dshape2vecset,chang2015shapenet,objaverse,objaverseXL}. 
Among all 3D generation methods, there is a line of work that utilizes diffusion models to generate 3DGS~\cite{mu2024gsd, roessle2024l3dg, zhang2024gaussiancube, zou2024triplane,yushigaussiananything,tang2024dreamgaussian}. 
The unstructured nature of 3DGS poses a significant challenge in finding a shared canonical space to train diffusion model. 
GSD~\cite{mu2024gsd} constraints the number of 3D Gaussians, while L3DG~\cite{roessle2024l3dg} embeds 3D Gaussians into a dense latent grid.
TriplaneGaussian~\cite{zou2024triplane} and DiffGS~\cite{zhou2024diffgs} directly decode 3D Gaussian attributes from generated Triplanes~\cite{chan2022efficient}. 
Omegas~\cite{yan2024omages64} trains a 2D diffusion model to predict 2D UV maps of the geometry and materials of 3D objects. 
To generate dynamic 3D objects, a recent method~\cite{ren2023dreamgaussian4d} explicitly introduces the time dimension by leveraging HexPlanes~\cite{cao2023hexplane}. 
Nevertheless, none of the existing methods is capable of generating articulated 3D humans due to their inherent complexity and large deformations. 
To address this, our method introduces the diffusion process in the network weight space, which encapsulates the information of dynamic human avatars. 
The concept of diffusion in network weight space has been explored in areas such as shape generation~\cite{erkocc2023hyperdiffusion} and transfer learning~\cite{soro2025diffusion}. However, our approach is the first to leverage hyper diffusion for dynamic human avatar generation.

\section{Background}
\begin{figure*}[ht]
  \centering
  \includegraphics[width=\linewidth]{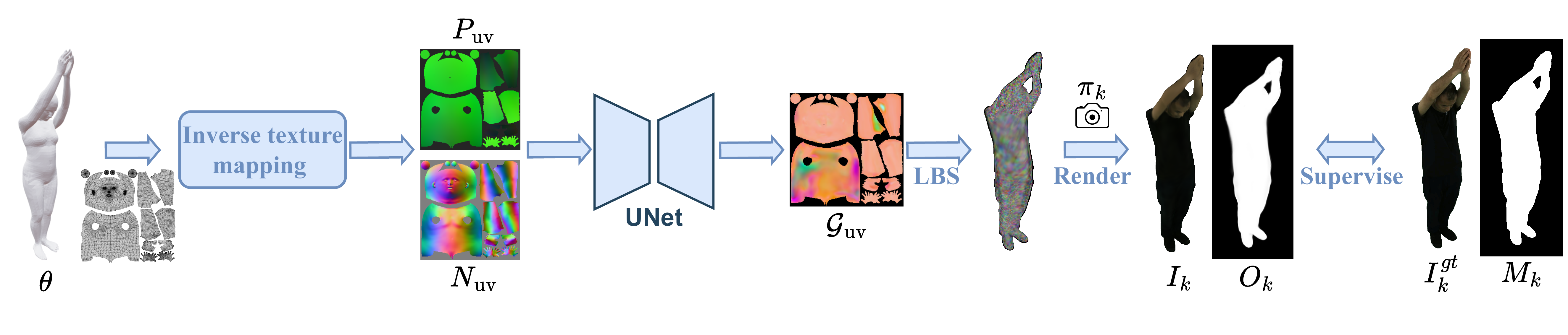}
  \caption{\textbf{Dynamic human representation learning based on UNet.} Given a specific human pose, the pose-dependent position and normal maps are generated via inverse texture mapping. These maps serve as inputs to the UNet, which predicts pose-dependent 3D Gaussians for rendering. During training, the UNet is optimized using multi-view RGB image sequences along with their corresponding segmentation masks.}
  \vspace{-4mm}
  \label{fig:overfit}
\end{figure*}

\subsection{SMPL-X} \label{subsec:smplx}
SMPL-X~\cite{pavlakos2019smplx} is a 3D parametric human model that represents the human shape (without cloth) consisting of body, hands, and face. 
This model consists of $10{,}475$ vertices and $54$ joints, allowing for control over body shape, body pose, and face expression. 
The deformation process can be defined as
\begin{equation}
    {\cal M}(\beta,\theta,\psi)={LBS}(T_{P}(\beta,\theta,\psi),J(\beta),\theta,{W}),
\end{equation}
where $\beta$, $\theta$ and $\psi$ represent shape, pose and expression parameters, respectively. 
The linear blend skinning (LBS) function~\cite{lewis2000pose}, denoted as ${LBS}(\cdot)$, is used to transform the canonical template $T_{P}$ to the given pose $\theta$ based on the skinning weight ${W}$ and joint locations $J(\beta)$. 
The canonical template $T_{P}$ can be computed as
\begin{equation}
    \label{eq:smpl_x}
    T_{P}(\beta,\theta,\psi)=T_{C}+B_{S}(\beta)+ B_{P}(\theta) + B_{E}(\psi),
\end{equation}
where $T_{C}$ is the mean shape and $B_{S}(\beta), B_{P}(\theta), B_{E}(\psi)$ represent per-vertex displacements calculated by the blend shapes ${S}, {P}, {E}$ with their corresponding shape, pose, expression parameters. 
\subsection{3D Gaussian Splatting} \label{subsec:3dgs}
3D Gaussian Splatting~\cite{kerbl20233d} is an explicit point-based representation for novel view synthesis and 3D reconstruction that models static scenes using a collection of 3D Gaussian primitives. 
These primitives enable real-time rendering through differentiable rasterization. Each Gaussian primitive is parametrized by their center position $\mu \in \mathbb{R}^{3}$, covariance $\Sigma \in \mathbb{R}^{3\times 3}$, color $c \in \mathbb{R}^{3}$, and opacity $\alpha \in \mathbb{R}$. 
By projecting 3D Gaussians onto the camera’s imaging plane, the 2D Gaussians are assigned and sorted to the corresponding tiles for point-based rendering~\cite{zwicker2002ewa}, i.e.
\begin{equation}
    \mathrm{c}(p)=\sum_{i\in N}c_{i}\sigma_{i}\prod_{j=1}^{i-1}(1-\sigma_{j}),
\end{equation}
where $\sigma_{i}=\alpha_{i}\exp(-\frac12(p-\mu_{i})^{\mathsf{T}}\Sigma_{i}^{-1}(p-\mu_{i})),$ and $p$ is the location of queried point and $\mu_i, \Sigma_i, c_i, \alpha_i$ and $\sigma_i$ are the center position, covariance, color, opacity, and density of the $i$-th Gaussian primitive, respectively. 
To model view-dependent appearance, the color $c$ is represented via coefficients of spherical harmonics (SH)~\cite{kerbl20233d}. 
In practice, each Gaussian is parametrized as $\mathcal{G}_i = \{{p}_i, {s}_i, {q}_i, {\alpha}_i, {h}_i\} \in \mathbb{R}^{59}$, including 3D center position ${p}_i \in \mathbb{R}^{3}$, scaling ${s}_i \in \mathbb{R}^{3}$, quaternion ${q}_i \in \mathbb{R}^{4}$, opacity ${\alpha}_i \in \mathbb{R}$, and spherical harmonics ${h}_i \in \mathbb{R}^{48}$.

\subsection{Denoising diffusion models} \label{subsec:diffu}
Given a dataset of examples drawn independently from a real data distribution $q({x})$, diffusion models aim to learn the data distribution by sequentially denoising random noise samples~\cite{ho2020denoising,sohl2015deep,song2019generative}.
During training, the diffusion model defines a forward diffusion process in which a small amount of Gaussian noise is added in $T$ steps, producing a sequence of noisy samples ${x}_{1},...,{x}_{T}$. 
The step sizes are controlled by a variance schedule $\{\beta_{t}\in(0,1)\}_{t=1}^{T}$, i.e.

\begin{align}
q({ x}_{t}|{x}_{t-1})&=N({x}_{t};\sqrt{1-\beta_{t}}{x}_{t-1},\beta_{t}{ I}), \\
q({x}_{1:T}|{x}_{0})&=\prod_{t=1}^{T}q({x}_{t}|{x}_{t-1}).
\end{align}

During inference, the reverse process iteratively removes noise from an input $x_{T}$ drawn from the Gaussian distribution using the learned denoiser and obtains a clean sample $x_0$ in the end~\cite{ho2020denoising,song2020denoising}.

\section{Our method} \label{sec:method}
In this work, we present an unconditional generative model for synthesizing dynamic human avatars trained on a large, cross-identity, and multi-view human video dataset. 
Our approach involves two stages. 
In the first stage, we train a UNet~\cite{ronneberger2015u} to map 3D skeletal human poses to the corresponding pose-dependent 3DGS for each human avatar individually. 
In the second stage, we propose a transformer-based~\cite{radford2019language} hyper diffusion model for generative and photorealistic human modeling, which is trained on the collection of network weights obtained from the first stage. 
At inference, our model can generate network weights corresponding to valid dynamic human avatars by performing the reverse diffusion process on randomly sampled noise.
\begin{figure*}[!ht]
  \centering
  \includegraphics[width=\linewidth]{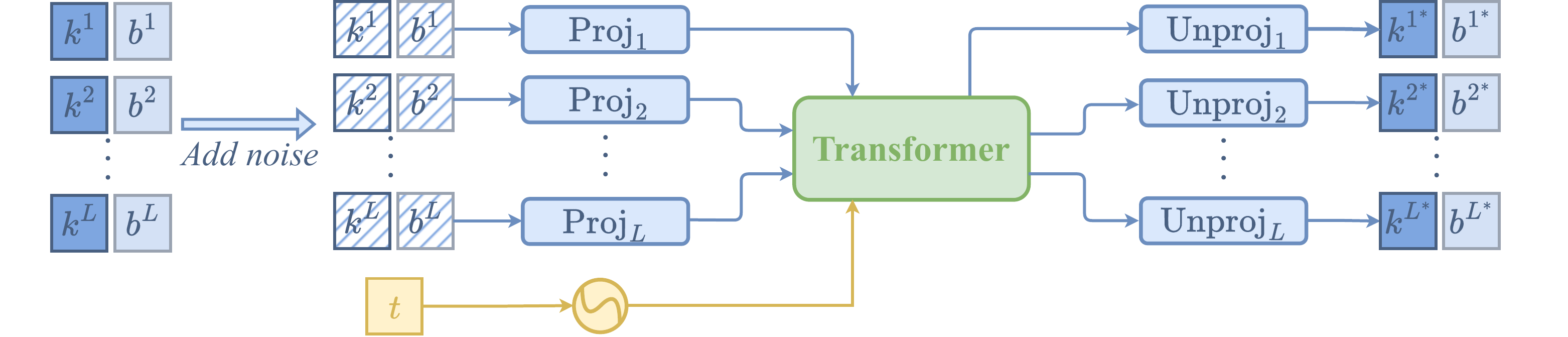}
  \caption{\textbf{Diffusion process on network weight space.} During the forward diffusion process, the standard Gaussian noise at time step $t$ is added to the network weights and the transformer take the noisy weights as well as the time step $t$ to predict the denoised weights.}
  \vspace{-4mm}
  \label{fig:diffusion}
\end{figure*}

\subsection{Dynamic human representation}
The overall pipeline of dynamic human representation learning is depicted in Fig.~\ref{fig:overfit}. Inspired by recent advances in person-specific dynamic human rendering~\cite{pang2024ash,li2024animatable,kwon2024generalizable}, we model each individual human avatar with a dedicated lightweight UNet with network weight $w$, denoted as $\mathcal{U}_{w}$.
The UNet takes pose-dependent texture as input, specifically the  normal texture $N_\mathrm{uv}(\theta) \in \mathbb{R}^{T\times T\times 3}$ and position texture $P_\mathrm{uv}(\theta) \in \mathbb{R}^{T\times T\times 3}$, which together encode the body pose $\theta$ in the 2D UV space. 
These textures are derived from the posed template $T_{P}$ (see Eq.\ref{eq:smpl_x}) via inverse texture mapping~\cite{pang2024ash}. 
Notably, unlike previous person-specific methods~\cite{pang2024ash,li2024animatable}, we utilize the mean SMPL-X template (i.e.\ $\beta = 0, \psi = 0$) instead of a person-specific template mesh, enabling a unified input motion representation across different individuals (i.e.\ shared UV space and mesh template). 
The UNet outputs 3D Gaussians parameterized in the same 2D UV space (i.e. $\mathcal{G}_\mathrm{uv}(\theta) \in \mathbb{R}^{T\times T\times 59} $), such that each texel of the template mesh encodes the parameters of a corresponding 3D Gaussian. This approach effectively binds the Gaussians to the template, enabling accurate and flexible avatar representation. 
To this end, the UNet learns the pose-dependent 3DGS, i.e.
\begin{equation}
    \mathcal{G}_\mathrm{uv}(\theta) = \mathcal{U}_{w}(N_\mathrm{uv}(\theta), P_\mathrm{uv}(\theta)).
\end{equation}
After obtaining the Gaussians, we use LBS~\cite{lewis2000pose} to transform the positions of Gaussians from canonical pose space to world space:
\begin{equation}
    \mathrm{}{p}_\mathrm{uv} = LBS((a_a \cdot \bar{v}_{a} + a_b \cdot \bar{v}_{b}+ a_c \cdot \bar{v}_{c})+{d}_\mathrm{uv}),
\end{equation}
where ${d}_\mathrm{uv} \in \mathbb{R}^{T\times T\times 3}$ denotes the learned offset of Gaussians, ${p}_\mathrm{uv} \in \mathbb{R}^{T\times T\times 3}$ represents the final positions of Gaussians in world space, $a_{\bullet}$ is the barycentric weight on each texel and $\bar{v}_{\bullet}$ is the corresponding canonical vertex position of the template mesh. To this end, our method models the pose-dependent deformations by learning the pose-dependent offset of Gaussians $d_{\mathrm{uv}}$.

For each camera view $k$ with projection matrix $\pi_k$, the resulting 3D Gaussians $\mathcal{G}_\mathrm{uv}$ are rendered using a differentiable Gaussian rasterizer $\mathcal{R}$, producing a RGB image $I_k \in \mathbb{R}^{H \times W \times 3}$ and an opacity image $O_k \in \mathbb{R}^{H \times W \times 1}$, i.e. 
\begin{equation}
    (I_k, O_k) = \mathcal{R}(\mathcal{G}_\mathrm{uv}(\theta), \pi_k).
\end{equation}
To train the UNet $\mathcal{U}_{w}$, we compute the mean absolute error $\mathcal{L}_{\mathrm{L1}}$ and the structural similarity $\mathcal{L}_{\mathrm{SSIM}}$ between the rendered RGB image $I_{k}$ and the ground-truth image $I_{k}^{gt}$, following prior works~\cite{pang2024ash,kerbl20233d}. 
Additionally, we compute the AlexNet-based~\cite{krizhevsky2012imagenet} perceptual loss~\cite{zhang2018unreasonable} $\mathcal{L}_{\mathrm{LPIPS}}$ for better visual appearance and the mean absolute error $\mathcal{L}_{\mathrm{mask}}$ between the rendered opacity image $O_{k}$ and the ground-truth human segmentation mask $M_{k}$ for better outlines of Gaussian primitives~\cite{cao2022authentic}. 
The overall training loss is a weighted sum of the individual losses, i.e.
\begin{equation}
\begin{split}
    \mathcal{L}_{\mathrm{total}} = 
    &\ \lambda_{\mathrm{pix}}\mathcal{L}_{\mathrm{L1}}(I_k, I_k^{gt})  
    + \lambda_{\mathrm{str}}\mathcal{L}_{\mathrm{SSIM}}(I_k, I_k^{gt}) + \\
    &  \lambda_{\mathrm{per}}\mathcal{L}_{\mathrm{LPIPS}}(I_k, I_k^{gt}) 
    + \lambda_{\mathrm{m}}\mathcal{L}_{\mathrm{mask}}(O_k, M_k).
\end{split}
\label{eq:loss}
\end{equation}
\par 
In this manner, each dynamic human avatar is represented by its corresponding neural network weights $w_i$, providing a unified canonical space that accommodates variations in shape and appearance across different individuals. 
Thus, the per-instance optimization leads to a collection of network weights $\mathcal{W} = \{w_i\}_{i=1}^{N}$, where $N$ is the number of human individuals. 
To constrain the network weight distribution, we use a consistent weight initialization~\cite{erkocc2023hyperdiffusion} instead of random weight initialization.

\subsection{Network weight space diffusion}
Once we obtain the collection of network weights $\mathcal{W}$, we train a diffusion model to learn the underlying distribution of the network weights as shown in Fig.~\ref{fig:diffusion}. 
We consider the set of weights of a given UNet $w_i$ as a sequence of convolutional kernels and biases, i.e.
\begin{equation}
    w_i = \{k_i^l, b_i^l\}_{l=1}^{L},
\end{equation}
where $k_i^l \in \mathbb{R}^{C^l_{\mathrm{out}}\times C^l_{\mathrm{in}}\times K_h^l \times K_w^l}$ and $b_i^l \in \mathbb{R}^{C_{\mathrm{out}}^{l}}$ are the kernel and bias of $l$-th convolutional layer, respectively. 
During the forward diffusion process, standard Gaussian noise at step $t$ is added to the network weights $w_i$ and we employ a transformer architecture $\mathcal{T}$ as our diffusion model, following recent approaches~\cite{erkocc2023hyperdiffusion,peebles2022learning}. 
Specifically, for each layer, we add Gaussian noise and flatten the kernel weights and concatenate the corresponding biases, treating this combined vector as a distinct token for the transformer input. This process partitions the entire set of weights $w_i$ into $L$ separate tokens, one for each layer. The layer-wise partitioning preserves the hierarchical structure of the network, which helps the transformer to more effectively capture and learn the complex network weight space. This is in contrast to previous methods~\cite{peebles2022learning,soro2025diffusion}, which flatten all network weights into a single 1D vector and chunk it into tokens, potentially losing important structural information.
Before passing these tokens to the transformer, we project each one into a shared feature space using a linear layer for each token, i.e.
\begin{equation}
    t_{\mathrm{in}}^{i} = \text{Proj}_{i}(k^i\oplus b^i), \text{ for } i=\{1, \ldots, L\},
\end{equation}
where $t_{in}^{i}$ is the projected input token and $\oplus$ indicates concatenation. As a result, tokens with the same dimension can be directly used as input for the transformer. 
The resulting noisy tokens, along with the sinusoidal embedding of $t$, are then fed into the transformer $\mathcal{T}$, i.e.
\begin{equation}
    (t_{\mathrm{out}}^{1}, t_{\mathrm{out}}^{2}, \ldots, t_{\mathrm{out}}^{L}) = \mathcal{T}(t_{\mathrm{in}}^{1}, t_{\mathrm{in}}^{2}, \ldots, t_{\mathrm{in}}^{L}, \text{emb}(t)).
\end{equation}
The transformer $\mathcal{T}$ consists of multiple self-attention and feed-forward layers, facilitating effective information exchange, both within and across the tokens representing each layers.
After processing through the transformer, we unproject each token back to its original dimension using a separate linear layer for each token, mirroring the projection performed at the input, i.e.
\begin{equation}
    (k^{i^{*}}\oplus b^{i^{*}}) = \text{Unproj}_{i}(t_{\mathrm{out}}^{i}), \text{ for } i=\{1, \ldots, L\}.
\end{equation}
This yields the denoised network weights $w^*$. 
\begin{figure*}[ht]
  \centering
  \includegraphics[width=\linewidth]{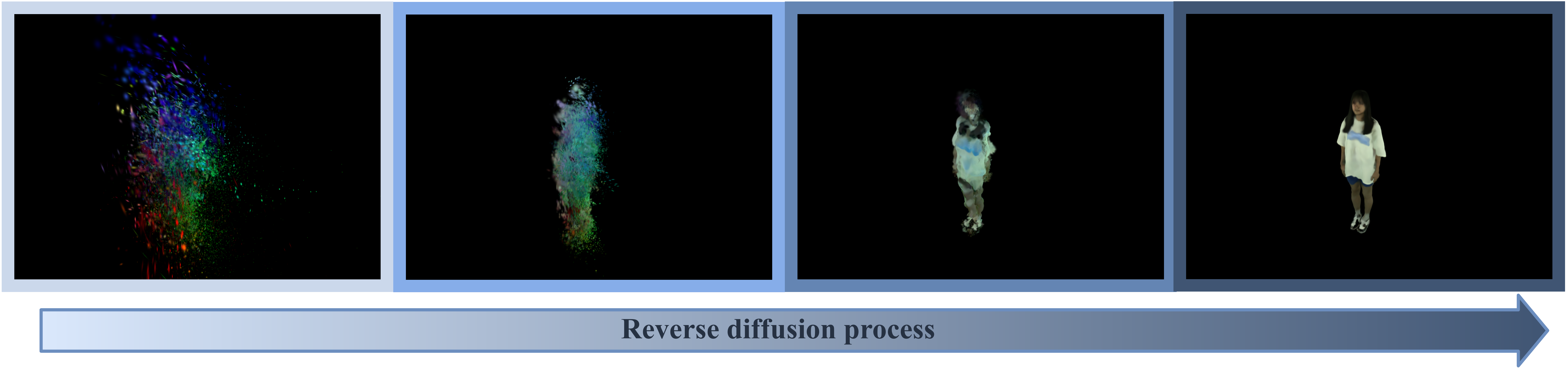}
  \caption{\textbf{Denoising network weights at various time steps.} Here the network weights are visualized based on the rendering images. The rendering images show the reverse diffusion process based on DDIM sampling. We observe that UNet weights corrupted by noise fail to represent a valid human avatar. However, the iterative denoising process yields a high-quality human avatar.}
  \label{fig:denoising}
  \vspace{-4mm}
\end{figure*}

Following prior approaches~\cite{ho2020denoising,erkocc2023hyperdiffusion}, we train the a Mean Squared Error (MSE) loss between the denoised weights $w^*$ and the input weights $w$. 
During inference we utilize DDIM~\cite{song2020denoising} to sample network weights from the diffusion model. 
Fig.~\ref{fig:denoising} shows an example of the denoising process to generate a valid UNet that represents a dynamic human avatar. 

\subsection{Implementation details}
Rather than using the original UNet~\cite{ronneberger2015u}, which contains approximately 30 million learnable parameters and thus poses significant challenges for the diffusion process, we empirically reduce the number of hidden channels to 64, resulting in a lightweight network with only 0.6 million parameters. 
Each avatar-specific UNet is trained using the AdamW optimizer~\cite{kingma2014adam} with a batch size of 1 and a learning rate of $1 \times 10^{-4}$. In terms of the training loss, we empirically set $\lambda_{\mathrm{pix}} = 1.0, \lambda_{\mathrm{str}} = 0.1, \lambda_{\mathrm{per}} = 0.01$, and $\lambda_{\mathrm{m}} = 0.1$ in Eq.\ref{eq:loss}. 
During training, the images are downsampled to $1024 \times 750$ and cropped using the segmentation mask, while the UV map resolution is set to $256$ for efficiency. 
Following prior work~\cite{pang2024ash}, we employ a 30k-step warm-up and train the UNet for a total of 700k iterations. 
In the context of network weight space diffusion, we utilize a transformer architecture comprising 12 blocks, each equipped with multi-head self-attention (16 heads) and a feed-forward layer with a hidden dimension of 2048. 
For training the diffusion model, network weights are standardized to zero mean and unit variance. 
We use the AdamW optimizer~\cite{kingma2014adam} with a batch size of 16 and a learning rate of $2 \times 10^{-4}$. 
The learning rate is reduced by $10\%$ every 200 epochs. 
We train the transformer for $6000$ epochs until convergence. 

\begin{figure*}[ht]
  \centering
  \includegraphics[width=\linewidth]{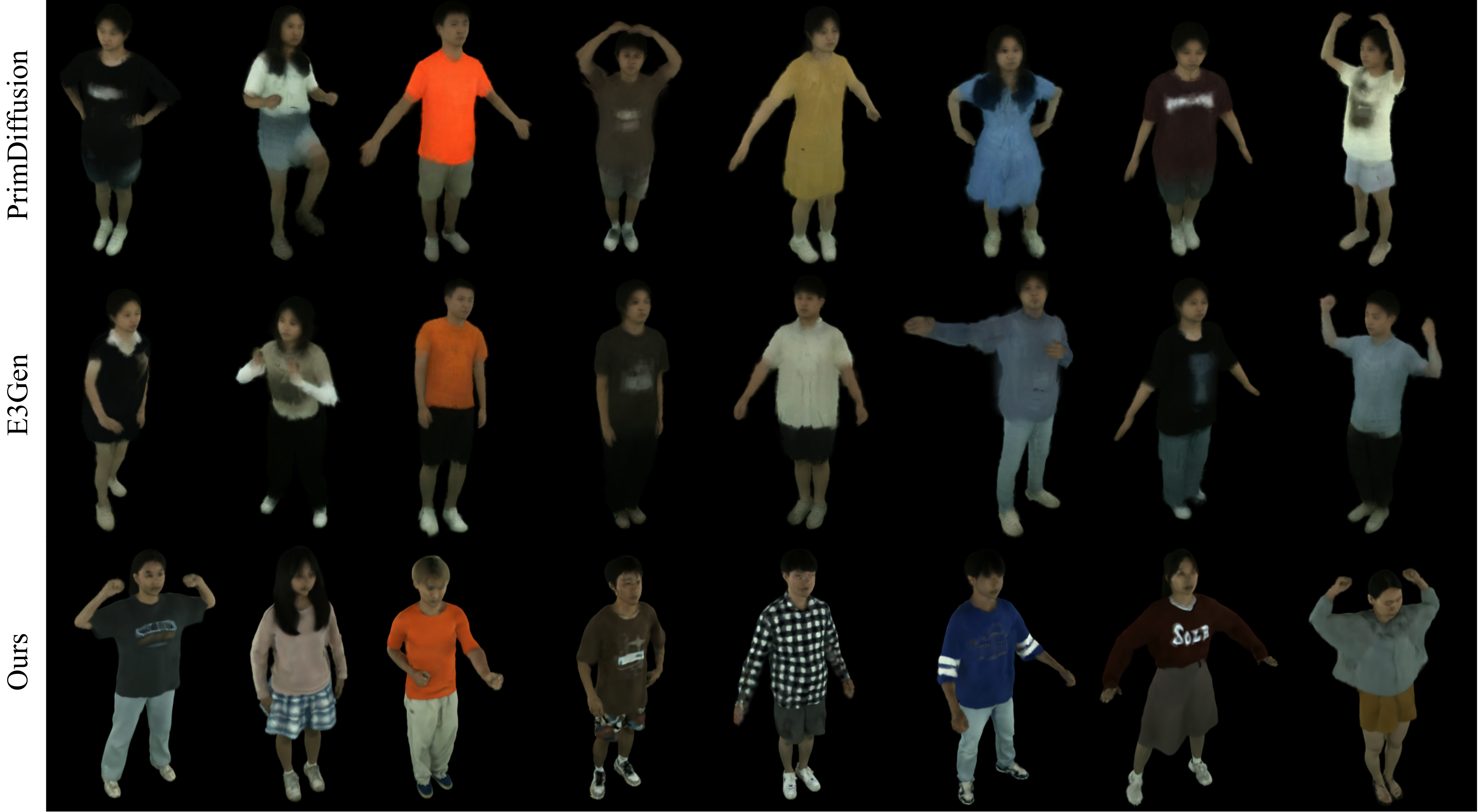}
  \caption{\textbf{Qualitative results on unconditional human avatar generation.} Compared to baseline methods, our method is able to generate more photorealistic human avatars.}
  \label{fig:comparison}
\end{figure*}

\section{Experimental results}
\label{sec:experiments}
\subsection{Datasets and metrics}
For evaluation, we utilize the multi-view human dataset MVHumanNet~\cite{xiong2024mvhumannet}, which contains a large number of diverse identities with everyday clothing. To be comparable with baseline methods~\cite{zhang20243gen,chen2023primdiffusion}, we manually select 500 video sequences from the first 1500 sequences based on the SMPL-X pose parameter estimation accuracy. Evaluation of unconditional generation of dynamic human avatar can be challenging due to the lack of direct correspondence to ground truth data~\cite{erkocc2023hyperdiffusion}. Following prior unconditional generation methods~\cite{yang2019pointflow,luo2021diffusion,erkocc2023hyperdiffusion}, we evaluate the methods based on Minimum Matching Distance (MMD), Coverage (COV), and 1-Nearest-Neighbor Accuracy (1-NNA), i.e.
\begin{align*}
\text{MMD}(S_g, S_r) &= \frac{1}{|S_r|} \sum_{Y \in S_r} \min_{X \in S_g} D(X, Y), \\
\text{COV}(S_g, S_r) &= \frac{|\{ \arg \min_{Y \in S_r} D(X, Y) \mid X \in S_g \}|}{|S_r|}, \\
\text{1-NNA}(S_g, S_r) &= \frac{\sum_{X \in S_g} \mathbf{1}[N_X \in S_g] + \sum_{Y \in S_r} \mathbf{1}[N_Y \in S_r]}{|S_g| + |S_r|},
\end{align*}
where $S_g, S_r$ are the set of generated data and reference data respectively, $D(X, Y)$ is the distance function between data sample $X$ and $Y$, in the 1-NNA metric $N_X$ is a data sample that is closest to $X$ in both generated and reference dataset, i.e.,
\begin{equation*}
    N_X = \mathrm{argmin}_{K\in S_g \cup S_r / X}D(X, K).
\end{equation*}

Here, we use the PSNR, and the LPIPS~\cite{zhang2018unreasonable} with AlexNet features~\cite{krizhevsky2012imagenet} (scaled by 1000) as the distance functions between the rendered images from generated human avatars and the corresponding ground-truth images. Following baseline methods~\cite{zhang20243gen,chen2023primdiffusion}, we also adopt Fréchet Inception Distance (FID)~\cite{heusel2017gans} and  Kernel Inception Distance (KID)~\cite{binkowski2018demystifying} to evaluate the quality of rendered images based on the Inception-V3 model~\cite{szegedy2016rethinking}. 

\begin{table*}[ht]
\centering
\begin{tabular}{lcccccc}
\hline
\textbf{Methods} & $\text{MMD}_{\mathrm{PSNR}}$ $\uparrow$ & $\text{MMD}_{\mathrm{LPIPS}}$ $\downarrow$ & $\text{COV}_{\mathrm{PSNR}} (\%)$ $\uparrow$  & $\text{1-NNA}_{\mathrm{PSNR}} (\%)$ $\downarrow$  & $\text{FID}$ $\downarrow$ & $\text{KID}$ $\downarrow$  \\
\hline
PrimDiffusion~\cite{chen2023primdiffusion}  & 22.23 & 26.45 & 52.3 & 26.8 & 41.97 & 328.46 \\
E3Gen~\cite{zhang20243gen}  & 21.14 & 32.28 & 58.2 & 21.3 & 32.17 & 284.31 \\
Ours  & \textbf{27.52} & \textbf{12.13} & \textbf{63.8}  &  \textbf{15.7} & \textbf{12.68} & \textbf{123.26} \\
\hline
\end{tabular}
\caption{{\textbf{Quantitative results on unconditional human avatar generation.} Our method outperforms the prior state-of-the-art methods in the context of rendering quality as well as generation diversity.}
}
\vspace{-3mm}
\label{tab:comparison}

\end{table*}

\begin{figure*}[ht]
  \centering
  \includegraphics[width=0.99\linewidth]{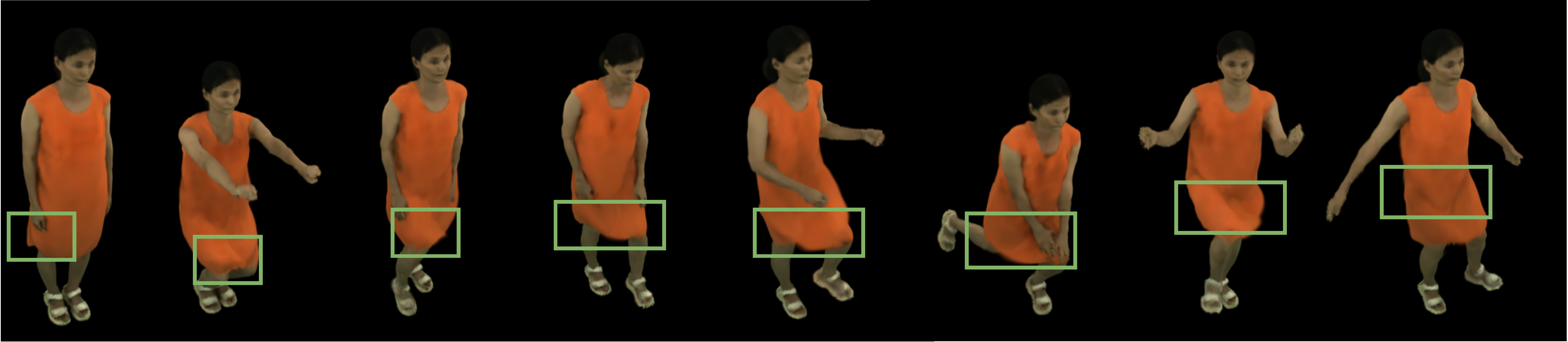}
  \caption{\textbf{An example of unconditional human avatar generation of our method.} Rendering sequence demonstrates our method's ability to generate dynamic human avatars. Pose-dependent deformations are emphasized with green rectangles.}
  \label{fig:dynamics}
\end{figure*}

\subsection{Comparison}
We compare our method to other human avatar generation methods training on multi-view human dataset: PrimDiffusion~\cite{chen2023primdiffusion}, E3Gen~\cite{zhang20243gen}. Notably, both methods can only generate static human avatars, which are animated based on simple skeleton-based articulation (i.e.~LBS~\cite{lewis2023pose}). We follow the same experiment settings to train them using the 500 multi-view human video sequences from MVHumanNet~\cite{xiong2024mvhumannet}. To evaluate unconditional generation performance, we generate 500 samples for each method. The quantitative results are summarized in Tab.~\ref{tab:comparison}. Compared to baseline methods, our approach generates more photorealistic renderings. Additionally, it outperforms existing techniques in terms of generation diversity. Fig.~\ref{fig:comparison} provides a qualitative comparison demonstrating our method’s ability to render more photorealistic dynamic human avatars. Moreover, our method is capable of generating dynamic human avatar with pose-dependent deformations as shown in Fig.~\ref{fig:dynamics}.

\section{Ablation study}
\label{sec:ablation}

\begin{table*}[!ht]
\centering
\begin{tabular}{lcccccc}
\hline
 & $\text{MMD}_{\mathrm{PSNR}}$ $\uparrow$ & $\text{MMD}_{\mathrm{LPIPS}}$ $\downarrow$ & $\text{COV}_{\mathrm{PSNR}} (\%)$ $\uparrow$  & $\text{1-NNA}_{\mathrm{PSNR}} (\%)$ $\downarrow$ & $\text{FID}$ $\downarrow$ & $\text{KID}$ $\downarrow$  \\
\hline
Latent diffusion & 21.23 & 27.28 & 0.4 & 98.0 & 58.52 & 480.34 \\
1D vector flatten & 27.12 & 12.30 & 54.2 & 27.6 & 14.73 & 134.65 \\
Ours  & \textbf{27.52} & \textbf{12.13} & \textbf{63.8}  &  \textbf{15.7} & \textbf{12.68} & \textbf{123.26} \\

\hline
\end{tabular}
\caption{{\textbf{Ablation study on the choice of the diffusion model.} Our layer-wise partition achieves the best performance in comparison to other choices.}
\vspace{-2mm}
}
\label{tab:ablation_model}
\end{table*}
In this section, we examine the different choices of the diffusion model. In contrast to images, which have a well-defined grid-like structure and can leverage specialized network architectures such as UNet~\cite{ronneberger2015u, rombach2022high} or DiT~\cite{peebles2023scalable} for diffusion models, network weights exhibit a more complex and less regular structure. As a result, selecting an appropriate representation for network weights, as well as an effective network architecture for the diffusion process, becomes crucial. Here we ablate different choices for network weight representation and network architectures to identify the most effective configuration. The experiment setting is the same as Sec.~\ref{sec:experiments}. Specifically, we compare our layer-wise partitioning to latent diffusion~\cite{rombach2022high} and 1D vector flatten.
\begin{figure}[ht]
  \centering
  \includegraphics[width=\linewidth]{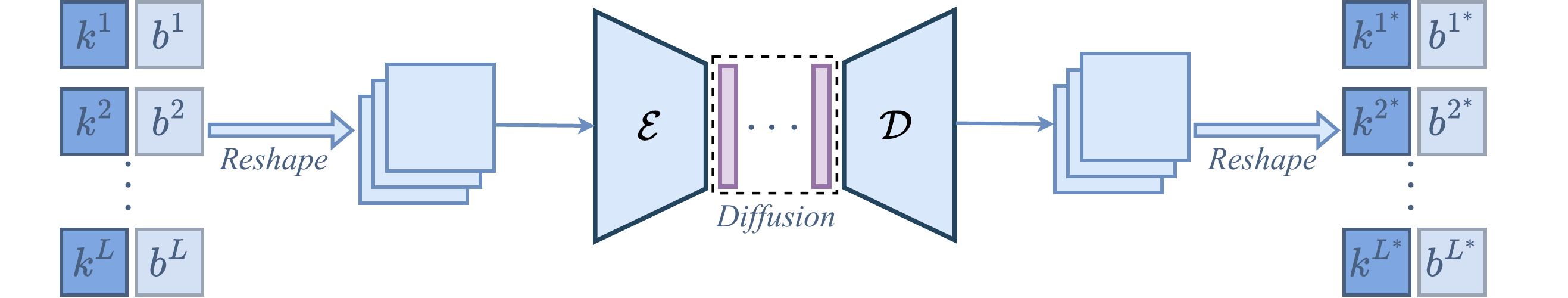}
  \caption{\textbf{Latent diffusion model on network weight space.} The network weights are first reshaped into a 2D feature map. An encoder then converts this 2D feature map into a latent space representation. The diffusion process takes place on the latent space. Afterward, a decoder transforms the latent features back into a 2D feature map. Finally, the network weights are recovered by reshaping the 2D feature map.}
  \vspace{-3mm}
  \label{fig:ldm}
\end{figure}
Fig.~\ref{fig:ldm} illustrates the process of latent diffusion on network weight space. Following recent work~\cite{soro2025diffusion}, the network weights are first reshaped into a 2D feature map. This feature map is then treated as an input image and processed using the standard latent diffusion model~\cite{rombach2022high}. To this end, the training contains two stages. In the first stage, the encoder and decoder are trained to reconstruct the input feature map, with a KL-penalty applied to encourage the latent features to follow a standard normal distribution. In the second stage, a diffusion model is trained on the latent space representation. In the context of 1D vector flattening, the network weights are first flattened into a single vector. This vector is then partitioned into $N_{\mathrm{chunk}}$ equal-sized chunks, each with dimension $C_{\mathrm{chunk}}$. If needed, zero-padding is used to ensure all chunks are the same size. Each chunk is subsequently treated as an input token for the transformer-based diffusion model.  Tab.~\ref{tab:ablation_model} summarizes the results of our ablation study and highlights that our layer-wise partition achieves the best performance. We observe that latent diffusion is prone to mode collapse, resulting in highly similar or nearly identical generations across samples.

\section{Limitation and future work}
\label{sec:limitation}
We introduce, for the first time, an unconditional generative model for synthesizing dynamic human avatars through network weight space diffusion. Unlike prior approaches that rely solely on simple skeleton-based articulation, our method enables the generation of photorealistic human avatars with complex, pose-dependent deformations. Despite these advancements, some limitations warrant further investigation. In the first training stage, our method optimizes a UNet to learn motion-aware 3D Gaussians. However, we observe that the UNet shows limited generalization to unseen poses, underscoring the need to enhance its ability to handle novel poses. Additionally, each human avatar is currently represented by a separate UNet, without addressing the entanglement between geometry and appearance. In future work, it would be valuable to explore methods for disentangling geometry and appearance by leveraging relationships across different human avatars~\cite{tewari2022disentangled3d}. In the context of hyper diffusion, the current method attempts to directly learn the complex, high-dimensional distribution of the network weight space, which poses significant challenges (e.g.~neural permutation symmetry~\cite{kunin2020neural,crisostomi2024c}) for training and limits generative performance. To address this, it would be valuable to explore approaches such as low-rank adaptation (e.g.~LoRA~\cite{hu2022lora}) or network basis learning~\cite{jaderberg2014speeding,bolager2023sampling}, which could simplify the learning process and enhance generation capabilities. 

\section{Conclusion}
In this work, we present the first method for dynamic human avatar generation that incorporates pose-dependent deformations. Our approach uniquely combines the strengths of person-specific rendering and diffusion-based generative modeling to achieve highly photorealistic results. Specifically, we optimize a set of UNets, each corresponding to an individual human avatar, and leverage a diffusion model trained over the network weights to enable avatar generation. Experimental results demonstrate that our method outperforms existing approaches by producing more photorealistic avatars with accurately learned pose-dependent deformations by evaluating on a large-scale, cross-identity, multi-view video dataset. This contribution paves the way for more realistic human avatar generation in a variety of applications.

\clearpage
{
    \small
    \bibliographystyle{ieeenat_fullname}
    \bibliography{main}

\begin{thebibliography}{99}
\providecommand{\natexlab}[1]{#1}
\providecommand{\url}[1]{\texttt{#1}}
\expandafter\ifx\csname urlstyle\endcsname\relax
  \providecommand{\doi}[1]{doi: #1}\else
  \providecommand{\doi}{doi: \begingroup \urlstyle{rm}\Url}\fi

\bibitem[Bergman et~al.(2022)Bergman, Kellnhofer, Yifan, Chan, Lindell, and Wetzstein]{bergman2022generative}
Alexander~W Bergman, Petr Kellnhofer, Wang Yifan, Eric~R Chan, David~B Lindell, and Gordon Wetzstein.
\newblock Generative neural articulated radiance fields.
\newblock In \emph{Advances in Neural Information Processing Systems}, 2022.

\bibitem[Bickel et~al.(2007)Bickel, Botsch, Angst, Matusik, Otaduy, Pfister, and Gross]{bickel2007multi}
Bernd Bickel, Mario Botsch, Roland Angst, Wojciech Matusik, Miguel Otaduy, Hanspeter Pfister, and Markus Gross.
\newblock Multi-scale capture of facial geometry and motion.
\newblock \emph{ACM Transactions on Graphics (ToG)}, 26\penalty0 (3):\penalty0 33--es, 2007.

\bibitem[Bi{\'n}kowski et~al.(2018)Bi{\'n}kowski, Sutherland, Arbel, and Gretton]{binkowski2018demystifying}
Miko{\l}aj Bi{\'n}kowski, Danica~J Sutherland, Michael Arbel, and Arthur Gretton.
\newblock Demystifying mmd gans.
\newblock In \emph{International Conference on Learning Representations}, 2018.

\bibitem[Bolager et~al.(2023)Bolager, Burak, Datar, Sun, and Dietrich]{bolager2023sampling}
Erik~L Bolager, Iryna Burak, Chinmay Datar, Qing Sun, and Felix Dietrich.
\newblock Sampling weights of deep neural networks.
\newblock In \emph{Advances in Neural Information Processing Systems}, 2023.

\bibitem[Cao and Johnson(2023)]{cao2023hexplane}
Ang Cao and Justin Johnson.
\newblock Hexplane: A fast representation for dynamic scenes.
\newblock In \emph{Proceedings of the IEEE/CVF Conference on Computer Vision and Pattern Recognition}, 2023.

\bibitem[Cao et~al.(2022)Cao, Simon, Kim, Schwartz, Zollhoefer, Saito, Lombardi, Wei, Belko, Yu, et~al.]{cao2022authentic}
Chen Cao, Tomas Simon, Jin~Kyu Kim, Gabe Schwartz, Michael Zollhoefer, Shun-Suke Saito, Stephen Lombardi, Shih-En Wei, Danielle Belko, Shoou-I Yu, et~al.
\newblock Authentic volumetric avatars from a phone scan.
\newblock \emph{ACM Transactions on Graphics (ToG)}, 41\penalty0 (4):\penalty0 1--19, 2022.

\bibitem[Cao et~al.(2024)Cao, Cao, Han, Shan, and Wong]{cao2024dreamavatar}
Yukang Cao, Yan-Pei Cao, Kai Han, Ying Shan, and Kwan-Yee~K Wong.
\newblock Dreamavatar: Text-and-shape guided 3d human avatar generation via diffusion models.
\newblock In \emph{Proceedings of the IEEE/CVF Conference on Computer Vision and Pattern Recognition}, 2024.

\bibitem[Chan et~al.(2022)Chan, Lin, Chan, Nagano, Pan, De~Mello, Gallo, Guibas, Tremblay, Khamis, et~al.]{chan2022efficient}
Eric~R Chan, Connor~Z Lin, Matthew~A Chan, Koki Nagano, Boxiao Pan, Shalini De~Mello, Orazio Gallo, Leonidas~J Guibas, Jonathan Tremblay, Sameh Khamis, et~al.
\newblock Efficient geometry-aware 3d generative adversarial networks.
\newblock In \emph{Proceedings of the IEEE/CVF conference on computer vision and pattern recognition}, 2022.

\bibitem[Chang et~al.(2015)Chang, Funkhouser, Guibas, Hanrahan, Huang, Li, Savarese, Savva, Song, Su, et~al.]{chang2015shapenet}
Angel~X Chang, Thomas Funkhouser, Leonidas Guibas, Pat Hanrahan, Qixing Huang, Zimo Li, Silvio Savarese, Manolis Savva, Shuran Song, Hao Su, et~al.
\newblock Shapenet: An information-rich 3d model repository.
\newblock \emph{arXiv preprint arXiv:1512.03012}, 2015.

\bibitem[Chen et~al.(2023)Chen, Hong, Mei, Wang, Yang, and Liu]{chen2023primdiffusion}
Zhaoxi Chen, Fangzhou Hong, Haiyi Mei, Guangcong Wang, Lei Yang, and Ziwei Liu.
\newblock Primdiffusion: Volumetric primitives diffusion for 3d human generation.
\newblock In \emph{Advances in Neural Information Processing Systems}, 2023.

\bibitem[Cordner and Fong(2023)]{lewis2023pose}
Matt Cordner and Nickson Fong.
\newblock Pose space deformation: a unified approach to shape interpolation and skeleton-driven deformation.
\newblock In \emph{Seminal Graphics Papers: Pushing the Boundaries, Volume 2}, pages 811--818. Association for Computing Machinery, 2023.

\bibitem[Crisostomi et~al.(2024)Crisostomi, Fumero, Baieri, Bernard, and Rodola]{crisostomi2024c}
Donato Crisostomi, Marco Fumero, Daniele Baieri, Florian Bernard, and Emanuele Rodola.
\newblock Cycle-consistent multi-model merging.
\newblock In \emph{Advances in Neural Information Processing Systems}, 2024.

\bibitem[Deitke et~al.(2022)Deitke, Schwenk, Salvador, Weihs, Michel, VanderBilt, Schmidt, Ehsani, Kembhavi, and Farhadi]{objaverse}
Matt Deitke, Dustin Schwenk, Jordi Salvador, Luca Weihs, Oscar Michel, Eli VanderBilt, Ludwig Schmidt, Kiana Ehsani, Aniruddha Kembhavi, and Ali Farhadi.
\newblock Objaverse: A universe of annotated 3d objects.
\newblock \emph{arXiv preprint arXiv:2212.08051}, 2022.

\bibitem[Deitke et~al.(2023)Deitke, Liu, Wallingford, Ngo, Michel, Kusupati, Fan, Laforte, Voleti, Gadre, VanderBilt, Kembhavi, Vondrick, Gkioxari, Ehsani, Schmidt, and Farhadi]{objaverseXL}
Matt Deitke, Ruoshi Liu, Matthew Wallingford, Huong Ngo, Oscar Michel, Aditya Kusupati, Alan Fan, Christian Laforte, Vikram Voleti, Samir~Yitzhak Gadre, Eli VanderBilt, Aniruddha Kembhavi, Carl Vondrick, Georgia Gkioxari, Kiana Ehsani, Ludwig Schmidt, and Ali Farhadi.
\newblock Objaverse-xl: A universe of 10m+ 3d objects.
\newblock \emph{arXiv preprint arXiv:2307.05663}, 2023.

\bibitem[Erko{\c{c}} et~al.(2023)Erko{\c{c}}, Ma, Shan, Nie{\ss}ner, and Dai]{erkocc2023hyperdiffusion}
Ziya Erko{\c{c}}, Fangchang Ma, Qi Shan, Matthias Nie{\ss}ner, and Angela Dai.
\newblock Hyperdiffusion: Generating implicit neural fields with weight-space diffusion.
\newblock In \emph{Proceedings of the IEEE/CVF international conference on computer vision}, 2023.

\bibitem[Esser et~al.(2024)Esser, Kulal, Blattmann, Entezari, M{\"u}ller, Saini, Levi, Lorenz, Sauer, Boesel, et~al.]{esser2024scaling}
Patrick Esser, Sumith Kulal, Andreas Blattmann, Rahim Entezari, Jonas M{\"u}ller, Harry Saini, Yam Levi, Dominik Lorenz, Axel Sauer, Frederic Boesel, et~al.
\newblock Scaling rectified flow transformers for high-resolution image synthesis.
\newblock In \emph{International Conference on Machine Learning}, 2024.

\bibitem[Feng et~al.(2022)Feng, Yang, Pollefeys, Black, and Bolkart]{feng2022capturing}
Yao Feng, Jinlong Yang, Marc Pollefeys, Michael~J Black, and Timo Bolkart.
\newblock Capturing and animation of body and clothing from monocular video.
\newblock In \emph{SIGGRAPH Asia}, 2022.

\bibitem[Garrido et~al.(2013)Garrido, Valgaerts, Wu, and Theobalt]{garrido2013reconstructing}
Pablo Garrido, Levi Valgaerts, Chenglei Wu, and Christian Theobalt.
\newblock Reconstructing detailed dynamic face geometry from monocular video.
\newblock \emph{ACM Transactions on Graphics (ToG)}, 32\penalty0 (6):\penalty0 158--1, 2013.

\bibitem[Habermann et~al.(2023)Habermann, Liu, Xu, Pons-Moll, Zollhoefer, and Theobalt]{habermann2023hdhumans}
Marc Habermann, Lingjie Liu, Weipeng Xu, Gerard Pons-Moll, Michael Zollhoefer, and Christian Theobalt.
\newblock Hdhumans: A hybrid approach for high-fidelity digital humans.
\newblock \emph{Proceedings of the ACM on Computer Graphics and Interactive Techniques}, 6\penalty0 (3):\penalty0 1--23, 2023.

\bibitem[Heusel et~al.(2017)Heusel, Ramsauer, Unterthiner, Nessler, and Hochreiter]{heusel2017gans}
Martin Heusel, Hubert Ramsauer, Thomas Unterthiner, Bernhard Nessler, and Sepp Hochreiter.
\newblock Gans trained by a two time-scale update rule converge to a local nash equilibrium.
\newblock In \emph{Advances in Neural Information Processing Systems}, 2017.

\bibitem[Ho et~al.(2020)Ho, Jain, and Abbeel]{ho2020denoising}
Jonathan Ho, Ajay Jain, and Pieter Abbeel.
\newblock Denoising diffusion probabilistic models.
\newblock In \emph{Advances in Neural Information Processing Systems}, 2020.

\bibitem[Hu et~al.(2022)Hu, Shen, Wallis, Allen-Zhu, Li, Wang, Wang, Chen, et~al.]{hu2022lora}
Edward~J Hu, Yelong Shen, Phillip Wallis, Zeyuan Allen-Zhu, Yuanzhi Li, Shean Wang, Lu Wang, Weizhu Chen, et~al.
\newblock Lora: Low-rank adaptation of large language models.
\newblock In \emph{International Conference on Learning Representations}, 2022.

\bibitem[Hu et~al.(2024{\natexlab{a}})Hu, Zhang, Zhang, Zhou, Liu, Zhang, and Nie]{hu2024gaussianavatar}
Liangxiao Hu, Hongwen Zhang, Yuxiang Zhang, Boyao Zhou, Boning Liu, Shengping Zhang, and Liqiang Nie.
\newblock Gaussianavatar: Towards realistic human avatar modeling from a single video via animatable 3d gaussians.
\newblock In \emph{Proceedings of the IEEE/CVF conference on computer vision and pattern recognition}, 2024{\natexlab{a}}.

\bibitem[Hu et~al.(2023)Hu, Hong, Pan, Mei, Yang, and Liu]{hu2023sherf}
Shoukang Hu, Fangzhou Hong, Liang Pan, Haiyi Mei, Lei Yang, and Ziwei Liu.
\newblock Sherf: Generalizable human nerf from a single image.
\newblock In \emph{Proceedings of the IEEE/CVF International Conference on Computer Vision}, 2023.

\bibitem[Hu et~al.(2024{\natexlab{b}})Hu, Hu, and Liu]{hu2024gauhuman}
Shoukang Hu, Tao Hu, and Ziwei Liu.
\newblock Gauhuman: Articulated gaussian splatting from monocular human videos.
\newblock In \emph{Proceedings of the IEEE/CVF conference on computer vision and pattern recognition}, 2024{\natexlab{b}}.

\bibitem[Huang et~al.(2024)Huang, Yi, Xiu, Liao, Tang, Cai, and Thies]{huang2024tech}
Yangyi Huang, Hongwei Yi, Yuliang Xiu, Tingting Liao, Jiaxiang Tang, Deng Cai, and Justus Thies.
\newblock Tech: Text-guided reconstruction of lifelike clothed humans.
\newblock In \emph{International Conference on 3D Vision (3DV)}, 2024.

\bibitem[Jaderberg et~al.(2014)Jaderberg, Vedaldi, and Zisserman]{jaderberg2014speeding}
Max Jaderberg, Andrea Vedaldi, and Andrew Zisserman.
\newblock Speeding up convolutional neural networks with low rank expansions.
\newblock \emph{arXiv preprint arXiv:1405.3866}, 2014.

\bibitem[Kerbl et~al.(2023)Kerbl, Kopanas, Leimk{\"u}hler, and Drettakis]{kerbl20233d}
Bernhard Kerbl, Georgios Kopanas, Thomas Leimk{\"u}hler, and George Drettakis.
\newblock 3d gaussian splatting for real-time radiance field rendering.
\newblock \emph{ACM Transactions on Graphics (ToG)}, 42\penalty0 (4):\penalty0 139--1, 2023.

\bibitem[Kingma and Ba(2014)]{kingma2014adam}
Diederik~P Kingma and Jimmy Ba.
\newblock Adam: A method for stochastic optimization.
\newblock \emph{arXiv preprint arXiv:1412.6980}, 2014.

\bibitem[Kolotouros et~al.(2023)Kolotouros, Alldieck, Zanfir, Bazavan, Fieraru, and Sminchisescu]{kolotouros2023dreamhuman}
Nikos Kolotouros, Thiemo Alldieck, Andrei Zanfir, Eduard Bazavan, Mihai Fieraru, and Cristian Sminchisescu.
\newblock Dreamhuman: Animatable 3d avatars from text.
\newblock In \emph{Advances in Neural Information Processing Systems}, 2023.

\bibitem[Krizhevsky et~al.(2012)Krizhevsky, Sutskever, and Hinton]{krizhevsky2012imagenet}
Alex Krizhevsky, Ilya Sutskever, and Geoffrey~E Hinton.
\newblock Imagenet classification with deep convolutional neural networks.
\newblock In \emph{Advances in Neural Information Processing Systems}, 2012.

\bibitem[Kunin et~al.(2020)Kunin, Sagastuy-Brena, Ganguli, Yamins, and Tanaka]{kunin2020neural}
Daniel Kunin, Javier Sagastuy-Brena, Surya Ganguli, Daniel~LK Yamins, and Hidenori Tanaka.
\newblock Neural mechanics: Symmetry and broken conservation laws in deep learning dynamics.
\newblock In \emph{International Conference on Learning Representations}, 2020.

\bibitem[Kwon et~al.(2023)Kwon, Liu, Fuchs, Habermann, and Theobalt]{kwon2023deliffas}
Youngjoong Kwon, Lingjie Liu, Henry Fuchs, Marc Habermann, and Christian Theobalt.
\newblock Deliffas: Deformable light fields for fast avatar synthesis.
\newblock In \emph{Advances in Neural Information Processing Systems}, 2023.

\bibitem[Kwon et~al.(2024)Kwon, Fang, Lu, Dong, Zhang, Carrasco, Mosella-Montoro, Xu, Takagi, Kim, et~al.]{kwon2024generalizable}
Youngjoong Kwon, Baole Fang, Yixing Lu, Haoye Dong, Cheng Zhang, Francisco~Vicente Carrasco, Albert Mosella-Montoro, Jianjin Xu, Shingo Takagi, Daeil Kim, et~al.
\newblock Generalizable human gaussians for sparse view synthesis.
\newblock In \emph{European Conference on Computer Vision}. Springer, 2024.

\bibitem[Lewis et~al.(2000)Lewis, Cordner, and Fong]{lewis2000pose}
JP Lewis, Matt Cordner, and Nickson Fong.
\newblock Pose space deformation: a unified approach to shape interpolation and skeleton-driven deformation.
\newblock In \emph{Proceedings of the 27th annual conference on Computer graphics and interactive techniques}, pages 165--172, 2000.

\bibitem[Li et~al.(2022)Li, Tanke, Vo, Zollh{\"o}fer, Gall, Kanazawa, and Lassner]{li2022tava}
Ruilong Li, Julian Tanke, Minh Vo, Michael Zollh{\"o}fer, J{\"u}rgen Gall, Angjoo Kanazawa, and Christoph Lassner.
\newblock Tava: Template-free animatable volumetric actors.
\newblock In \emph{European Conference on Computer Vision}, 2022.

\bibitem[Li et~al.(2024)Li, Zheng, Wang, and Liu]{li2024animatable}
Zhe Li, Zerong Zheng, Lizhen Wang, and Yebin Liu.
\newblock Animatable gaussians: Learning pose-dependent gaussian maps for high-fidelity human avatar modeling.
\newblock In \emph{Proceedings of the IEEE/CVF conference on computer vision and pattern recognition}, 2024.

\bibitem[Liao et~al.(2024)Liao, Yi, Xiu, Tang, Huang, Thies, and Black]{liao2024tada}
Tingting Liao, Hongwei Yi, Yuliang Xiu, Jiaxiang Tang, Yangyi Huang, Justus Thies, and Michael~J Black.
\newblock Tada! text to animatable digital avatars.
\newblock In \emph{International Conference on 3D Vision (3DV)}, 2024.

\bibitem[Lipman et~al.(2022)Lipman, Chen, Ben-Hamu, Nickel, and Le]{lipman2022flow}
Yaron Lipman, Ricky~TQ Chen, Heli Ben-Hamu, Maximilian Nickel, and Matthew Le.
\newblock Flow matching for generative modeling.
\newblock In \emph{International Conference on Learning Representations}, 2022.

\bibitem[Liu et~al.(2021)Liu, Habermann, Rudnev, Sarkar, Gu, and Theobalt]{liu2021neural}
Lingjie Liu, Marc Habermann, Viktor Rudnev, Kripasindhu Sarkar, Jiatao Gu, and Christian Theobalt.
\newblock Neural actor: Neural free-view synthesis of human actors with pose control.
\newblock \emph{ACM Transactions on Graphics (ToG)}, 40\penalty0 (6):\penalty0 1--16, 2021.

\bibitem[Liu et~al.(2024)Liu, Zhan, Tang, Shan, Zeng, Lin, Liu, and Liu]{liu2024humangaussian}
Xian Liu, Xiaohang Zhan, Jiaxiang Tang, Ying Shan, Gang Zeng, Dahua Lin, Xihui Liu, and Ziwei Liu.
\newblock Humangaussian: Text-driven 3d human generation with gaussian splatting.
\newblock In \emph{Proceedings of the IEEE/CVF Conference on Computer Vision and Pattern Recognition}, 2024.

\bibitem[Lombardi et~al.(2021)Lombardi, Simon, Schwartz, Zollhoefer, Sheikh, and Saragih]{lombardi2021mixture}
Stephen Lombardi, Tomas Simon, Gabriel Schwartz, Michael Zollhoefer, Yaser Sheikh, and Jason Saragih.
\newblock Mixture of volumetric primitives for efficient neural rendering.
\newblock \emph{ACM Transactions on Graphics (ToG)}, 40\penalty0 (4):\penalty0 1--13, 2021.

\bibitem[Long et~al.(2024)Long, Guo, Lin, Liu, Dou, Liu, Ma, Zhang, Habermann, Theobalt, et~al.]{long2024wonder3d}
Xiaoxiao Long, Yuan-Chen Guo, Cheng Lin, Yuan Liu, Zhiyang Dou, Lingjie Liu, Yuexin Ma, Song-Hai Zhang, Marc Habermann, Christian Theobalt, et~al.
\newblock Wonder3d: Single image to 3d using cross-domain diffusion.
\newblock In \emph{Proceedings of the IEEE/CVF conference on computer vision and pattern recognition}, 2024.

\bibitem[Loper et~al.(2015)Loper, Mahmood, Romero, Pons-Moll, and Black]{loper2015smpl}
Matthew Loper, Naureen Mahmood, Javier Romero, Gerard Pons-Moll, and Michael~J Black.
\newblock Smpl: A skinned multi-person linear model.
\newblock \emph{ACM Transactions on Graphics (ToG)}, 34\penalty0 (6), 2015.

\bibitem[Luo and Hu(2021)]{luo2021diffusion}
Shitong Luo and Wei Hu.
\newblock Diffusion probabilistic models for 3d point cloud generation.
\newblock In \emph{Proceedings of the IEEE/CVF conference on computer vision and pattern recognition}, 2021.

\bibitem[Meng et~al.(2023)Meng, Rombach, Gao, Kingma, Ermon, Ho, and Salimans]{meng2023distillation}
Chenlin Meng, Robin Rombach, Ruiqi Gao, Diederik Kingma, Stefano Ermon, Jonathan Ho, and Tim Salimans.
\newblock On distillation of guided diffusion models.
\newblock In \emph{Proceedings of the IEEE/CVF Conference on Computer Vision and Pattern Recognition}, 2023.

\bibitem[Metzer et~al.(2023)Metzer, Richardson, Patashnik, Giryes, and Cohen-Or]{metzer2023latent}
Gal Metzer, Elad Richardson, Or Patashnik, Raja Giryes, and Daniel Cohen-Or.
\newblock Latent-nerf for shape-guided generation of 3d shapes and textures.
\newblock In \emph{Proceedings of the IEEE/CVF conference on computer vision and pattern recognition}, 2023.

\bibitem[Mildenhall et~al.(2021)Mildenhall, Srinivasan, Tancik, Barron, Ramamoorthi, and Ng]{mildenhall2021nerf}
Ben Mildenhall, Pratul~P Srinivasan, Matthew Tancik, Jonathan~T Barron, Ravi Ramamoorthi, and Ren Ng.
\newblock Nerf: Representing scenes as neural radiance fields for view synthesis.
\newblock \emph{Communications of the ACM}, 65\penalty0 (1):\penalty0 99--106, 2021.

\bibitem[Mu et~al.(2024)Mu, Zuo, Guo, Wang, Lu, Wu, Xu, Dai, Yan, and Cheng]{mu2024gsd}
Yuxuan Mu, Xinxin Zuo, Chuan Guo, Yilin Wang, Juwei Lu, Xiaofeng Wu, Songcen Xu, Peng Dai, Youliang Yan, and Li Cheng.
\newblock Gsd: View-guided gaussian splatting diffusion for 3d reconstruction.
\newblock In \emph{European Conference on Computer Vision}, 2024.

\bibitem[Pang et~al.(2024)Pang, Zhu, Kortylewski, Theobalt, and Habermann]{pang2024ash}
Haokai Pang, Heming Zhu, Adam Kortylewski, Christian Theobalt, and Marc Habermann.
\newblock Ash: Animatable gaussian splats for efficient and photoreal human rendering.
\newblock In \emph{Proceedings of the IEEE/CVF Conference on Computer Vision and Pattern Recognition}, 2024.

\bibitem[Pavlakos et~al.(2019)Pavlakos, Choutas, Ghorbani, Bolkart, Osman, Tzionas, and Black]{pavlakos2019smplx}
Georgios Pavlakos, Vasileios Choutas, Nima Ghorbani, Timo Bolkart, Ahmed A.~A. Osman, Dimitrios Tzionas, and Michael~J. Black.
\newblock Expressive body capture: {3D} hands, face, and body from a single image.
\newblock In \emph{Proceedings of the IEEE/CVF international conference on computer vision}, 2019.

\bibitem[Peebles and Xie(2023)]{peebles2023scalable}
William Peebles and Saining Xie.
\newblock Scalable diffusion models with transformers.
\newblock In \emph{Proceedings of the IEEE/CVF international conference on computer vision}, 2023.

\bibitem[Peebles et~al.(2022)Peebles, Radosavovic, Brooks, Efros, and Malik]{peebles2022learning}
William Peebles, Ilija Radosavovic, Tim Brooks, Alexei~A Efros, and Jitendra Malik.
\newblock Learning to learn with generative models of neural network checkpoints.
\newblock \emph{arXiv preprint arXiv:2209.12892}, 2022.

\bibitem[Podell et~al.(2023)Podell, English, Lacey, Blattmann, Dockhorn, M{\"u}ller, Penna, and Rombach]{podell2023sdxl}
Dustin Podell, Zion English, Kyle Lacey, Andreas Blattmann, Tim Dockhorn, Jonas M{\"u}ller, Joe Penna, and Robin Rombach.
\newblock Sdxl: Improving latent diffusion models for high-resolution image synthesis.
\newblock \emph{arXiv preprint arXiv:2307.01952}, 2023.

\bibitem[Poole et~al.(2022)Poole, Jain, Barron, and Mildenhall]{poole2022dreamfusion}
Ben Poole, Ajay Jain, Jonathan~T Barron, and Ben Mildenhall.
\newblock Dreamfusion: Text-to-3d using 2d diffusion.
\newblock In \emph{International Conference on Learning Representations}, 2022.

\bibitem[Qian et~al.(2024)Qian, Mai, Hamdi, Ren, Siarohin, Li, Lee, Skorokhodov, Wonka, Tulyakov, and Ghanem]{qian2023magic123}
Guocheng Qian, Jinjie Mai, Abdullah Hamdi, Jian Ren, Aliaksandr Siarohin, Bing Li, Hsin-Ying Lee, Ivan Skorokhodov, Peter Wonka, Sergey Tulyakov, and Bernard Ghanem.
\newblock Magic123: One image to high-quality 3d object generation using both 2d and 3d diffusion priors.
\newblock In \emph{International Conference on Learning Representations}, 2024.

\bibitem[Radford et~al.(2019)Radford, Wu, Child, Luan, Amodei, Sutskever, et~al.]{radford2019language}
Alec Radford, Jeffrey Wu, Rewon Child, David Luan, Dario Amodei, Ilya Sutskever, et~al.
\newblock Language models are unsupervised multitask learners.
\newblock \emph{OpenAI blog}, 1\penalty0 (8):\penalty0 9, 2019.

\bibitem[Remelli et~al.(2022)Remelli, Bagautdinov, Saito, Wu, Simon, Wei, Guo, Cao, Prada, Saragih, et~al.]{remelli2022drivable}
Edoardo Remelli, Timur Bagautdinov, Shunsuke Saito, Chenglei Wu, Tomas Simon, Shih-En Wei, Kaiwen Guo, Zhe Cao, Fabian Prada, Jason Saragih, et~al.
\newblock Drivable volumetric avatars using texel-aligned features.
\newblock In \emph{ACM SIGGRAPH 2022 Conference Proceedings}, pages 1--9, 2022.

\bibitem[Ren et~al.(2023)Ren, Pan, Tang, Zhang, Cao, Zeng, and Liu]{ren2023dreamgaussian4d}
Jiawei Ren, Liang Pan, Jiaxiang Tang, Chi Zhang, Ang Cao, Gang Zeng, and Ziwei Liu.
\newblock Dreamgaussian4d: Generative 4d gaussian splatting.
\newblock \emph{arXiv preprint arXiv:2312.17142}, 2023.

\bibitem[Roessle et~al.(2024)Roessle, M{\"u}ller, Porzi, Rota~Bul{\`o}, Kontschieder, Dai, and Nie{\ss}ner]{roessle2024l3dg}
Barbara Roessle, Norman M{\"u}ller, Lorenzo Porzi, Samuel Rota~Bul{\`o}, Peter Kontschieder, Angela Dai, and Matthias Nie{\ss}ner.
\newblock L3dg: Latent 3d gaussian diffusion.
\newblock In \emph{SIGGRAPH Asia 2024 Conference Papers}, pages 1--11, 2024.

\bibitem[Rombach et~al.(2022)Rombach, Blattmann, Lorenz, Esser, and Ommer]{rombach2022high}
Robin Rombach, Andreas Blattmann, Dominik Lorenz, Patrick Esser, and Bj{\"o}rn Ommer.
\newblock High-resolution image synthesis with latent diffusion models.
\newblock In \emph{Proceedings of the IEEE/CVF conference on computer vision and pattern recognition}, 2022.

\bibitem[Ronneberger et~al.(2015)Ronneberger, Fischer, and Brox]{ronneberger2015u}
Olaf Ronneberger, Philipp Fischer, and Thomas Brox.
\newblock U-net: Convolutional networks for biomedical image segmentation.
\newblock In \emph{Medical image computing and computer-assisted intervention}. Springer, 2015.

\bibitem[Saito et~al.(2024)Saito, Schwartz, Simon, Li, and Nam]{saito2024relightable}
Shunsuke Saito, Gabriel Schwartz, Tomas Simon, Junxuan Li, and Giljoo Nam.
\newblock Relightable gaussian codec avatars.
\newblock In \emph{Proceedings of the IEEE/CVF conference on computer vision and pattern recognition}, 2024.

\bibitem[Sauer et~al.(2024)Sauer, Lorenz, Blattmann, and Rombach]{sauer2024adversarial}
Axel Sauer, Dominik Lorenz, Andreas Blattmann, and Robin Rombach.
\newblock Adversarial diffusion distillation.
\newblock In \emph{European Conference on Computer Vision}, 2024.

\bibitem[Shetty et~al.(2024)Shetty, Habermann, Sun, Luvizon, Golyanik, and Theobalt]{Shetty_2024_CVPR}
Ashwath Shetty, Marc Habermann, Guoxing Sun, Diogo Luvizon, Vladislav Golyanik, and Christian Theobalt.
\newblock Holoported characters: Real-time free-viewpoint rendering of humans from sparse rgb cameras.
\newblock In \emph{Proceedings of the IEEE/CVF international conference on computer vision}, 2024.

\bibitem[Shi et~al.(2023)Shi, Chen, Zhang, Liu, Xu, Wei, Chen, Zeng, and Su]{shi2023zero123++}
Ruoxi Shi, Hansheng Chen, Zhuoyang Zhang, Minghua Liu, Chao Xu, Xinyue Wei, Linghao Chen, Chong Zeng, and Hao Su.
\newblock Zero123++: a single image to consistent multi-view diffusion base model.
\newblock \emph{arXiv preprint arXiv:2310.15110}, 2023.

\bibitem[Sohl-Dickstein et~al.(2015)Sohl-Dickstein, Weiss, Maheswaranathan, and Ganguli]{sohl2015deep}
Jascha Sohl-Dickstein, Eric Weiss, Niru Maheswaranathan, and Surya Ganguli.
\newblock Deep unsupervised learning using nonequilibrium thermodynamics.
\newblock In \emph{International Conference on Machine Learning}, 2015.

\bibitem[Song et~al.(2020)Song, Meng, and Ermon]{song2020denoising}
Jiaming Song, Chenlin Meng, and Stefano Ermon.
\newblock Denoising diffusion implicit models.
\newblock In \emph{International Conference on Learning Representations}, 2020.

\bibitem[Song and Ermon(2019)]{song2019generative}
Yang Song and Stefano Ermon.
\newblock Generative modeling by estimating gradients of the data distribution.
\newblock In \emph{Advances in Neural Information Processing Systems}, 2019.

\bibitem[Soro et~al.(2025)Soro, Andreis, Lee, Jeong, Chong, Hutter, and Hwang]{soro2025diffusion}
Bedionita Soro, Bruno Andreis, Hayeon Lee, Wonyong Jeong, Song Chong, Frank Hutter, and Sung~Ju Hwang.
\newblock Diffusion-based neural network weights generation.
\newblock In \emph{International Conference on Learning Representations}, 2025.

\bibitem[Sun et~al.(2025)Sun, Dabral, Zhu, Fua, Theobalt, and Habermann]{sun2025real}
Guoxing Sun, Rishabh Dabral, Heming Zhu, Pascal Fua, Christian Theobalt, and Marc Habermann.
\newblock Real-time free-view human rendering from sparse-view rgb videos using double unprojected textures.
\newblock In \emph{Proceedings of the IEEE/CVF international conference on computer vision}, 2025.

\bibitem[Szegedy et~al.(2016)Szegedy, Vanhoucke, Ioffe, Shlens, and Wojna]{szegedy2016rethinking}
Christian Szegedy, Vincent Vanhoucke, Sergey Ioffe, Jon Shlens, and Zbigniew Wojna.
\newblock Rethinking the inception architecture for computer vision.
\newblock In \emph{Proceedings of the IEEE/CVF international conference on computer vision}, 2016.

\bibitem[Tang et~al.(2023)Tang, Wang, Zhang, Zhang, Yi, Ma, and Chen]{tang2023make}
Junshu Tang, Tengfei Wang, Bo Zhang, Ting Zhang, Ran Yi, Lizhuang Ma, and Dong Chen.
\newblock Make-it-3d: High-fidelity 3d creation from a single image with diffusion prior.
\newblock In \emph{Proceedings of the IEEE/CVF international conference on computer vision}, 2023.

\bibitem[Tang et~al.(2024)Tang, Ren, Zhou, Liu, and Zeng]{tang2024dreamgaussian}
Jiaxiang Tang, Jiawei Ren, Hang Zhou, Ziwei Liu, and Gang Zeng.
\newblock Dreamgaussian: Generative gaussian splatting for efficient 3d content creation.
\newblock In \emph{International Conference on Learning Representations}, 2024.

\bibitem[Teotia et~al.(2024)Teotia, Kim, Garrido, Habermann, Elgharib, and Theobalt]{teotia2024gaussianheads}
Kartik Teotia, Hyeongwoo Kim, Pablo Garrido, Marc Habermann, Mohamed Elgharib, and Christian Theobalt.
\newblock Gaussianheads: End-to-end learning of drivable gaussian head avatars from coarse-to-fine representations.
\newblock \emph{ACM Transactions on Graphics (ToG)}, 43\penalty0 (6):\penalty0 1--12, 2024.

\bibitem[Tewari et~al.(2022)Tewari, Pan, Fried, Agrawala, and Theobalt]{tewari2022disentangled3d}
Ayush Tewari, Xingang Pan, Ohad Fried, Maneesh Agrawala, and Christian Theobalt.
\newblock Disentangled3d: Learning a 3d generative model with disentangled geometry and appearance from monocular images.
\newblock In \emph{Proceedings of the IEEE/CVF conference on computer vision and pattern recognition}, 2022.

\bibitem[\url{https://renderpeople.com/3d-people/}(2018)]{renderpeople}
\url{https://renderpeople.com/3d-people/}.
\newblock Renderpeople, 2018.

\bibitem[Wang et~al.(2023{\natexlab{a}})Wang, Du, Li, Yeh, and Shakhnarovich]{wang2023score}
Haochen Wang, Xiaodan Du, Jiahao Li, Raymond~A Yeh, and Greg Shakhnarovich.
\newblock Score jacobian chaining: Lifting pretrained 2d diffusion models for 3d generation.
\newblock In \emph{Proceedings of the IEEE/CVF conference on computer vision and pattern recognition}, 2023{\natexlab{a}}.

\bibitem[Wang et~al.(2021)Wang, Liu, Liu, Theobalt, Komura, and Wang]{wang2021neus}
Peng Wang, Lingjie Liu, Yuan Liu, Christian Theobalt, Taku Komura, and Wenping Wang.
\newblock Neus: Learning neural implicit surfaces by volume rendering for multi-view reconstruction.
\newblock In \emph{Advances in Neural Information Processing Systems}, 2021.

\bibitem[Wang et~al.(2023{\natexlab{b}})Wang, Han, Habermann, Daniilidis, Theobalt, and Liu]{wang2023neus2}
Yiming Wang, Qin Han, Marc Habermann, Kostas Daniilidis, Christian Theobalt, and Lingjie Liu.
\newblock Neus2: Fast learning of neural implicit surfaces for multi-view reconstruction.
\newblock In \emph{Proceedings of the IEEE/CVF International Conference on Computer Vision (ICCV)}, 2023{\natexlab{b}}.

\bibitem[Weng et~al.(2022)Weng, Curless, Srinivasan, Barron, and Kemelmacher-Shlizerman]{weng2022humannerf}
Chung-Yi Weng, Brian Curless, Pratul~P Srinivasan, Jonathan~T Barron, and Ira Kemelmacher-Shlizerman.
\newblock Humannerf: Free-viewpoint rendering of moving people from monocular video.
\newblock In \emph{Proceedings of the IEEE/CVF conference on computer vision and pattern Recognition}, 2022.

\bibitem[Xiao et~al.(2025)Xiao, Zhang, Nie, Zhu, and Zheng]{RoGSplat2025CVPR}
Junjin Xiao, Qing Zhang, Yongwei Nie, Lei Zhu, and Wei-Shi Zheng.
\newblock {RoGSplat}: Learning robust generalizable human gaussian splatting from sparse multi-view images.
\newblock In \emph{Proceedings of the IEEE/CVF international conference on computer vision}, 2025.

\bibitem[Xiong et~al.(2024)Xiong, Li, Liu, Liao, Hu, Zhu, Ning, Qiu, Wang, Wang, et~al.]{xiong2024mvhumannet}
Zhangyang Xiong, Chenghong Li, Kenkun Liu, Hongjie Liao, Jianqiao Hu, Junyi Zhu, Shuliang Ning, Lingteng Qiu, Chongjie Wang, Shijie Wang, et~al.
\newblock Mvhumannet: A large-scale dataset of multi-view daily dressing human captures.
\newblock In \emph{Proceedings of the IEEE/CVF Conference on Computer Vision and Pattern Recognition}, 2024.

\bibitem[Xu et~al.(2023)Xu, Jiang, Wang, Fan, Wang, and Wang]{xu2023neurallift}
Dejia Xu, Yifan Jiang, Peihao Wang, Zhiwen Fan, Yi Wang, and Zhangyang Wang.
\newblock Neurallift-360: Lifting an in-the-wild 2d photo to a 3d object with 360deg views.
\newblock In \emph{Proceedings of the IEEE/CVF Conference on Computer Vision and Pattern Recognition}, 2023.

\bibitem[Yan et~al.(2024)Yan, Lee, Wan, and Chang]{yan2024omages64}
Xingguang Yan, Han-Hung Lee, Ziyu Wan, and Angel~X. Chang.
\newblock An object is worth 64x64 pixels: Generating 3d object via image diffusion, 2024.

\bibitem[Yang et~al.(2019)Yang, Huang, Hao, Liu, Belongie, and Hariharan]{yang2019pointflow}
Guandao Yang, Xun Huang, Zekun Hao, Ming-Yu Liu, Serge Belongie, and Bharath Hariharan.
\newblock Pointflow: 3d point cloud generation with continuous normalizing flows.
\newblock In \emph{Proceedings of the IEEE/CVF international conference on computer vision}, 2019.

\bibitem[Yu et~al.(2021)Yu, Zheng, Guo, Liu, Dai, and Liu]{tao2021function4d}
Tao Yu, Zerong Zheng, Kaiwen Guo, Pengpeng Liu, Qionghai Dai, and Yebin Liu.
\newblock Function4d: Real-time human volumetric capture from very sparse consumer rgbd sensors.
\newblock In \emph{Proceedings of the IEEE/CVF conference on computer vision and pattern recognition}, 2021.

\bibitem[Yushi et~al.(2025)Yushi, Zhou, Lyu, Hong, Yang, Dai, Pan, and Loy]{yushigaussiananything}
LAN Yushi, Shangchen Zhou, Zhaoyang Lyu, Fangzhou Hong, Shuai Yang, Bo Dai, Xingang Pan, and Chen~Change Loy.
\newblock Gaussiananything: Interactive point cloud flow matching for 3d generation.
\newblock In \emph{International Conference on Learning Representations}, 2025.

\bibitem[Zhang et~al.(2023)Zhang, Tang, Niessner, and Wonka]{zhang20233dshape2vecset}
Biao Zhang, Jiapeng Tang, Matthias Niessner, and Peter Wonka.
\newblock 3dshape2vecset: A 3d shape representation for neural fields and generative diffusion models.
\newblock \emph{ACM Transactions on Graphics (ToG)}, 42\penalty0 (4):\penalty0 1--16, 2023.

\bibitem[Zhang et~al.(2024{\natexlab{a}})Zhang, Cheng, Yang, Wang, Zhao, Tang, Chen, and Guo]{zhang2024gaussiancube}
Bowen Zhang, Yiji Cheng, Jiaolong Yang, Chunyu Wang, Feng Zhao, Yansong Tang, Dong Chen, and Baining Guo.
\newblock Gaussiancube: Structuring gaussian splatting using optimal transport for 3d generative modeling, 2024{\natexlab{a}}.

\bibitem[Zhang et~al.(2024{\natexlab{b}})Zhang, Li, Zhang, Cao, Shan, and Liao]{zhang2024humanref}
Jingbo Zhang, Xiaoyu Li, Qi Zhang, Yanpei Cao, Ying Shan, and Jing Liao.
\newblock Humanref: Single image to 3d human generation via reference-guided diffusion.
\newblock In \emph{Proceedings of the IEEE/CVF Conference on Computer Vision and Pattern Recognition}, 2024{\natexlab{b}}.

\bibitem[Zhang et~al.(2018)Zhang, Isola, Efros, Shechtman, and Wang]{zhang2018unreasonable}
Richard Zhang, Phillip Isola, Alexei~A Efros, Eli Shechtman, and Oliver Wang.
\newblock The unreasonable effectiveness of deep features as a perceptual metric.
\newblock In \emph{Proceedings of the IEEE conference on computer vision and pattern recognition}, 2018.

\bibitem[Zhang et~al.(2024{\natexlab{c}})Zhang, Yan, Liu, Sheng, and Yang]{zhang20243gen}
Weitian Zhang, Yichao Yan, Yunhui Liu, Xingdong Sheng, and Xiaokang Yang.
\newblock E 3gen: Efficient, expressive and editable avatars generation.
\newblock In \emph{Proceedings of the 32nd ACM International Conference on Multimedia}, 2024{\natexlab{c}}.

\bibitem[Zheng et~al.(2024)Zheng, Zhou, Shao, Liu, Zhang, Nie, and Liu]{zheng2024gps}
Shunyuan Zheng, Boyao Zhou, Ruizhi Shao, Boning Liu, Shengping Zhang, Liqiang Nie, and Yebin Liu.
\newblock Gps-gaussian: Generalizable pixel-wise 3d gaussian splatting for real-time human novel view synthesis.
\newblock In \emph{Proceedings of the IEEE/CVF conference on computer vision and pattern recognition}, 2024.

\bibitem[Zhou et~al.(2024)Zhou, Zhang, and Liu]{zhou2024diffgs}
Junsheng Zhou, Weiqi Zhang, and Yu-Shen Liu.
\newblock Diffgs: Functional gaussian splatting diffusion.
\newblock In \emph{Advances in Neural Information Processing Systems}, 2024.

\bibitem[Zhu et~al.(2024)Zhu, Zhan, Theobalt, and Habermann]{zhu2024trihuman}
Heming Zhu, Fangneng Zhan, Christian Theobalt, and Marc Habermann.
\newblock Trihuman: a real-time and controllable tri-plane representation for detailed human geometry and appearance synthesis.
\newblock \emph{ACM Transactions on Graphics (ToG)}, 44\penalty0 (1):\penalty0 1--17, 2024.

\bibitem[Zou et~al.(2024)Zou, Yu, Guo, Li, Liang, Cao, and Zhang]{zou2024triplane}
Zi-Xin Zou, Zhipeng Yu, Yuan-Chen Guo, Yangguang Li, Ding Liang, Yan-Pei Cao, and Song-Hai Zhang.
\newblock Triplane meets gaussian splatting: Fast and generalizable single-view 3d reconstruction with transformers.
\newblock In \emph{Proceedings of the IEEE/CVF conference on computer vision and pattern recognition}, 2024.

\bibitem[Zubekhin et~al.(2025)Zubekhin, Zhu, Gotardo, Beeler, Habermann, and Theobalt]{zubekhin2025giga}
Anton Zubekhin, Heming Zhu, Paulo Gotardo, Thabo Beeler, Marc Habermann, and Christian Theobalt.
\newblock Giga: Generalizable sparse image-driven gaussian avatars.
\newblock \emph{arXiv}, 2025.

\bibitem[Zwicker et~al.(2002)Zwicker, Pfister, Van~Baar, and Gross]{zwicker2002ewa}
Matthias Zwicker, Hanspeter Pfister, Jeroen Van~Baar, and Markus Gross.
\newblock Ewa splatting.
\newblock \emph{IEEE Transactions on Visualization and Computer Graphics}, 8\penalty0 (3):\penalty0 223--238, 2002.

\end{thebibliography}
}

\end{document}